\documentclass[prb, aps, twocolumn, groupedaddress,showpacs]{revtex4}
\usepackage[french,english]{babel} 
\usepackage{amsmath,amsfonts,amssymb}
\usepackage{graphicx}
\usepackage{float}
\usepackage{color}
\usepackage{version}

\begin{document}

\title{Spin-dependent Klein tunneling in polariton graphene with photonic spin-orbit interaction}

\author{Dmitry Solnyshkov}
\author{Anton Nalitov}
\author{Berihu Teklu}
\author{Louis Franck}
\author{Guillaume Malpuech}

\affiliation{Institut Pascal, PHOTON-N2, Universit\'e Clermont Auvergne, Blaise Pascal University, CNRS, 4 avenue Blaise Pascal, 63178 Aubi\`ere Cedex, France}

\begin{abstract}
We study Klein tunneling in polariton graphene. We show that the photonic spin-orbit coupling associated with the energy splitting between TE and TM photonic modes can be described as an emergent gauge field. It suppresses the Klein tunnelling in small energy range close to the Dirac points. Thanks to polariton spin-anisotropic interactions, polarized optical pumping allows to create potential barriers acting on a single polariton spin. We show that the resulting spin-dependent Klein tunneling can be used to create a perfectly transmitting polarization rotator operating at microscopic scale.
\end{abstract}
\maketitle

Emergent physics\cite{VolovikReview} in solid-state systems with effective Hamiltonians mimicking the behavior of less accessible systems has become a very productive field of research. The work of Semenoff \cite{Semenoff} was a starting point for the research on the analogs of effective electrodynamics in solids, which has culminated with beautiful works on graphene\cite{Katsnelson2006}, whose practical properties, such as the carrier mobility, are very much affected by the effect predicted in high-energy physics, but inaccessible there - Klein tunneling \cite{Klein}, leading to perfect transmission through a potential barrier by particle-antiparticle conversion.

Although graphene is a very promising, rich in effects \cite{CastroNeto2009,Novoselov2012}, and popular system, offering wide possibilities for the study of emergent physics \cite{Volovik2013,Volovik2014}, it has its own limitations: many measurements are indirect, its structure is fixed and its parameters (such as the band gap) are difficult to tune \cite{Zhou2007}. This is where different types of artificial graphene\cite{Polini2013} come into play. They can be based on different particles: atoms, both fermionic \cite{Tarruell2012,Jotzu2014} and bosonic \cite{Soltan2011,Duca2015}, or on photons and photonic quasi-particles \cite{Peleg2007,Kuhl2010,Kalesaki2014,Plotnik2014}, confined in a 2D honeycomb potential. At the lowest level of approximation, the single particle Hamiltonian is similar with the one of graphene, and it typically results in the same type of dispersion, characterized by the presence of the famous Dirac cones. These different systems offer a very wide tunability. The lattice parameters can be modified, specific types of spin-orbit interaction can be implemented \cite{Khanikaev2013,Goldman2010}. The physics of both bosonic and fermionic interacting systems can be addressed. In photonic systems, time-dependent perturbations can be used to create Floquet topological insulators \cite{Rechtsman2013}. Photonic systems in general allow a unique direct access to the time and spatial evolution of the wave functions (for example, the Bloch functions of a lattice in real and reciprocal space), with very simple experimental means \cite{Tanese,Tanese2014}. 

Recently, a 2D honeycomb lattice has been implemented for interacting photons, the exciton-polaritons (polaritons) \cite{Jacqmin2014}. These quasiparticles appear in microcavities\cite{Microcavities} in the strong coupling regime between the quantum well excitons and the cavity photons \cite{Weisbuch1992}. They combine light effective mass with strong interactions.  Their bosonic character provides the possibility for Bose-Einstein condensation \cite{Kasprzak2006}, while the two spin projections make it possible to describe them within the pseudospin formalism \cite{Whitlock1963} where the pseudospin dynamics is described by its coupling to effective magnetic fields \cite{Shelykh2010}. The spin-anisotropic character of the interactions \cite{Takemura2014} together with various controllable effective fields  offer a large variety of spin (polarization) effects for spin-optronics. The fabricated polariton graphene is based on a lattice of coupled micropillars \cite{Jacqmin2014}. This system is characterized by a spin-orbit interaction (SOI) acting on the real polarization of the photonic eigenstates. This SOI is induced by the energy splitting between the TE and TM polarized eigenmodes. It makes polariton graphene suitable for the realization of the optical spin Hall effect \cite{Nalitov2014}. As noticed in different contexts, the specific angular dependence of the TE-TM induced SOI induces chirality, which can generate stationary photonic spin currents \cite{Sala2014,Bliokh2015}. When combined with a Zeeman effective field in polariton graphene, it leads to the formation of a polaritonic analog of a Z topological insulator \cite{Liew2014,Nalitov2014b}.

\begin{figure}[t]\label{fig1}
\includegraphics[scale=0.35]{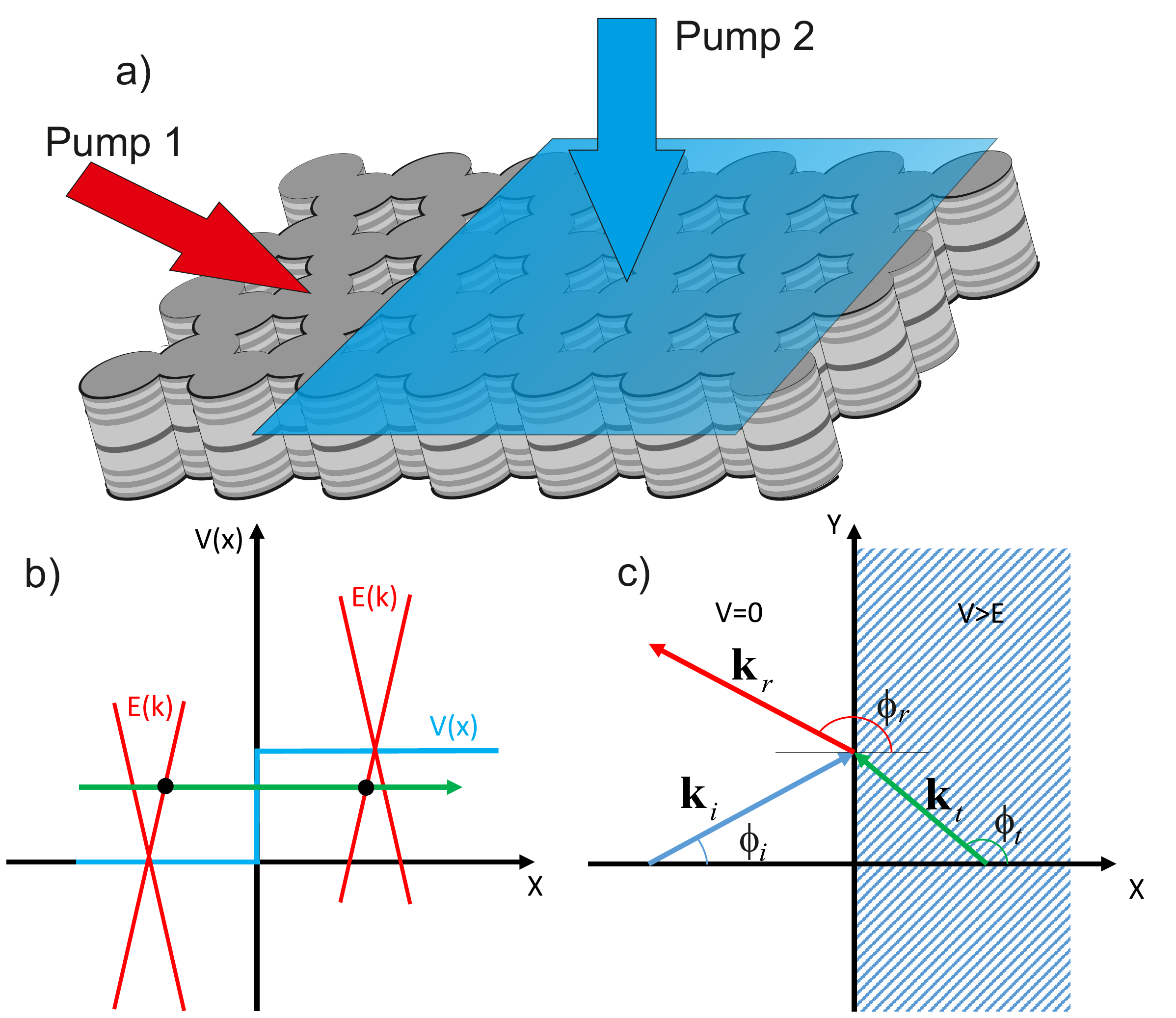} 
\caption{(color online) a) Scheme of the experiment. Patterned microcavity shown in grey, pump 1 (red) creates propagating polariton, pump 2 (blue) controls the barrier; b) Particle-antiparticle conversion in Klein tunneling: the dispersion is shown in red, the potential in blue, and the particle in green; c) Scheme of the Klein tunneling configuration. The indices $i$,$r$,$t$ stand for incident, reflected and transmitted waves. The barrier is shown by blue hatching.}
\end{figure} 

In this work, we propose and analyze an experimental scheme to study Klein tunneling in polariton graphene. The scheme of the proposed experiment is shown on Fig. 1(a). The potential barrier is created by a non-resonant optical pumping (blue), which populates an excitonic reservoir. This excitonic reservoir creates via the exciton-exciton interactions a potential barrier \cite{Wertz2010} affecting polaritons. The useful signal consists of polaritons optically injected close to the Dirac point of the dispersion by quasi-resonant pumping (red). The height of the barrier can be very precisely, controlled as demonstrated experimentally \cite{Sturm2013, HaiSon2013, Anton2013}.  Our study will include the role of the specific SOI present in the system, and also the freedom offered by the spin-anisotropic interactions between polaritons, which allows to optically create a potential barrier acting only on one specific spin component \cite{Amo2010}. Klein tunneling in the presence of Rashba like spin-orbit coupling has already been studied both for electrons \cite{Liu2012} and Bose-Einstein condensates \cite{Zhang2012}. Qualitatively similarly with their findings, we show that the TE-TM induced SOI considerably suppress the Klein tunneling in a narrow energy range around the Dirac point. We show that this field can be represented as an emergent gauge field. In the second part of the manuscript we propose a scheme of "Klein polarization rotator" which allows to optically control the polarization emitted by a the polariton graphene: if a potential barrier is created only for one of the two spin components, it affects the phase of this component, and thus the orientation of the linear polarization, while the transmission remains perfect due to the Klein tunneling.

\section{Tight-binding description of polariton graphene}
Before addressing Klein tunneling, we need to discuss the system, which provides the effective Hamiltonian required to achieve this phenomenon. Polariton graphene inherits the most important properties of real graphene. The simplest tight-binding model of electronic states in graphene \cite{Wallace} neglects the spin degree of freedom. Taking into account only the nearest-neighbor tunneling described by a constant $J$, the Hamiltonian in the basis of the two atoms $A$ and $B$ of the unit cell forming the lattice writes as follows:
\begin{equation}
\label{scalarHam}
{{\bf{H}}_{\bf{k}}} = \left( {\begin{array}{*{20}{c}}
0&{ - J{f_{\bf{k}}}}\\
{ - Jf_{\bf{k}}^\dag }&0
\end{array}} \right)
\end{equation}
where $f_{\mathbf{k}}=\sum_{j=1}^3 \exp(-\mathrm{i}\mathbf{k d}_{\varphi_j})$ is a sum over the 3 nearest atoms. This model gives rise to an emergent Dirac equation for electrons at the special points of the dispersion (called K and K' or simply Dirac points) located at the corners of the Brillouin zone. In this equation, the emergent "spin" 1/2 corresponds, in fact, to the sublattice degree of freedom -- atoms $A$ and $B$ (the coordinates in the reciprocal space are usually modified for convenience with respect to the full Hamiltonian as $k_x\rightarrow k_y$,$k_y\rightarrow -k_x$):
\begin{equation}
\label{scalarDirac}
\hat{H} = \hbar v_{F} \mathbf{k} . \hat{\mathbf{\sigma}} = \hbar v_{F} \begin{pmatrix}
0 & k_x - i k_y \\ k_x + i k_y & 0
\end{pmatrix}
\end{equation}
where $v_F=3Ja/2\hbar$ is the Fermi velocity, replacing the light speed in the original Dirac equation. For a bosonic particle, it should be considered simply as a parameter of the dispersion. The index of the dispersion branch is defined as $\alpha=\mathrm{sign} (E_{kin})=\pm 1$, $E_{kin}$ being the kinetic energy with its zero given by the energy position of the Dirac points. By the analogy with the Dirac equation, excitations with the branch index $\pm 1$  are called "particles" and "anti-particles" respectively.
In order to apply this Hamiltonian for exciton-polaritons, formed by strong coupling of two dimensional excitons and photons placed in a honeycomb potential, we need to make several approximations. The transverse (lateral) dynamics of photons in a cavity can be described using the Schrodinger equation. Restricting the consideration only to the lower polariton branch and using the parabolic approximation, valid close to the bottom of this branch, it becomes possible to apply the tight-binding approach, provided that the lattice sites are not too small (to avoid large wavevectors) and the confinement is strong enough, so that the bandwidth is smaller than the energy difference between the two first energy states of an isolated lattice site.

Periodic potential acting on photons and polaritons out of planar microcavities can be realized in different ways \cite{Kim2013,Cerda2010,Kim2014,Zhang2015}, the main example being however based on a lattice of micro-pillars obtained by patterning of a planar microcavity for which the above mentioned approximation is well realized.

\subsection{Spin-orbit coupling in polariton graphene}

Since we are going to deal with the polarization degree of freedom, we need to rewrite the tight-binding formalism, accounting for the two spin projections of polaritons, corresponding to right- and left-circular polarized photons. For this, we will have to work with bispinors instead of spinors. The corresponding tight-binding Hamiltonian for polaritons has been derived in \cite{Nalitov2014, Nalitov2014b}, together with extra terms responsible for the spin-orbit coupling, which will be discussed below. In the absence of spin-orbit coupling, the two circular components are completely independent, and the Hamiltonian is simply a combination of two tight-binding graphene Hamiltonians:

\begin{equation}
H_{\mathbf{k}}  = \left( {\begin{array}{*{20}c}
   0 & 0 & {Jf_{\mathbf{k}} } & 0  \\
   0 & 0 & 0 & {Jf_{\mathbf{k}} }  \\
   {Jf_{\mathbf{k}}^* } & 0 & 0 & 0  \\
   0 & {Jf_{\mathbf{k}}^* } & 0 & 0  \\

 \end{array} } \right)
\end{equation}
written in the basis $\Phi = \left( \Psi_A^+, \Psi_A^-, \Psi_B^+, \Psi_B^- \right)^{\mathrm{T}}$, with $\Psi_{A(B)}^\pm$ -- the wavefunctions of the two sublattices and two spin components. 

Although we are using the wavefunction description of polaritons, based on the Schrodinger equation, which is possible due to the quantization of photons in the microcavity in the growth direction, one should not forget that polaritons are formed from photons, described by the Maxwell's equations. It is natural to treat the electromagnetic waves on the basis of TE and TM eigenmodes, which are split in energy in presence of different media. For polaritons, the TE-TM splitting has been thoroughly discussed since Ref. \cite{Panzarini1999}, and many interesting effects based on this splitting have been demonstrated (e.g. optical spin Hall effect \cite{Kavokin2005,Leyder2007}, or acceleration of emergent magnetic monopoles \cite{Hivet}). In confined structures, the TE-TM splitting can be enhanced \cite{Dasbach2005} with respect to the planar cavities. For polariton graphene, it has already been the subject of extended experimental and theoretical studies \cite{Sala2015,Nalitov2014,Nalitov2014b}. Therefore, we need to take it into account in our description of Klein tunneling in polariton graphene. The Hamiltonian including the TE-TM induced spin-orbit coupling can be written as:

\begin{equation}\label{Ham_H}
\mathrm{H}_\mathbf{k} = \left( \begin{matrix}
0 & \mathrm{F}_{\mathbf{k}} \\
\mathrm{F}_{\mathbf{k}}^\dagger & 0
\end{matrix} \right),
\end{equation}
where the block matrices are defined as:

\begin{equation} \label{HamiltonianF}
\mathrm{F}_{\mathbf{k}} = - \left( \begin{matrix}
f_{\mathbf{k}} J & f_{\mathbf{k}}^+ \delta J \\
f_{\mathbf{k}}^- \delta J & f_{\mathbf{k}} J
\end{matrix} \right),
\end{equation}
Here, $J$ is the polarization-independent tunneling coefficient, whereas $\delta J$ is the SOI-induced polarization dependent term, which can be up to 10\% of $J$. Physically, it means that the polariton pseudospin rotates around the effective field during the tunneling process. The wavevector-dependent complex coefficients $f_{\mathbf{k}}$,$f_{\mathbf{k}}^\pm$  are defined by the sum over the nearest neighbors:
\begin{equation}
f_{\mathbf{k}}=\sum_{j=1}^3 \exp(-\mathrm{i}\mathbf{k d}_{\varphi_j}),\quad
f_{\mathbf{k}}^\pm = \sum_{j=1}^3 \exp(-\mathrm{i}\left[\mathbf{k d}_{\varphi_j} \mp 2 \varphi_j \right]), \notag
\end{equation}

Since the Klein tunneling occurs in the region of the reciprocal space close to the Dirac point, we can approximate the expressions for $f_{\mathbf{k}}$,$f_{\mathbf{k}}^\pm$, keeping only linear in $k$ terms around this point, which leads to the following expression: $\hat{H} = \hbar v_F \left(\hat{\sigma}_x\hat{k}_x+\hat{\sigma}_y\hat{k}_y \right)+\Delta\left(\hat{\sigma}_y\hat{s}_y -\hat{\sigma}_x\hat{s}_x\right)$, where $\Delta=3\delta J/2$ is the measure of the spin-orbit coupling, $\hat{s}_x$ and $\hat{s}_y$ being the pseudospin operators acting on the real polarization of particles. Without the loss of generality, we consider only one of the two Dirac points.   The matrix form of the Hamiltonian in this approximation can be written as:
\begin{widetext}
\begin{equation}
H = \begin{pmatrix}
0 & 0 & \hbar v_F ( k_x -i k_y ) & 2 \Delta \\
0 & 0 & 0 & \hbar v_F ( k_x -i k_y ) \\
\hbar v_F ( k_x +i k_y ) & 0 & 0 & 0 \\
2 \Delta & \hbar v_F (k_x + i k_y) & 0 & 0
\end{pmatrix}
\end{equation}
\end{widetext}

The 4 branches of the dispersion described by this Hamiltonian are parabolic for low wavevectors because of the spin-orbit coupling, and two of them are split-off by $\Delta$.

\begin{equation}
E = \pm \Delta \pm \sqrt{ \Delta^2 + (\hbar v_F k)^2}
\end{equation}

This result is the consequence of our approximation, valid only at intermediate values of wavevector $k$ relative to the Dirac points. In fact, the spin-orbit coupling in its full form leads to the trigonal warping of the dispersion. However, the typical scale of the trigonal warping is so small that it can be safely neglected \cite{Nalitov2014}. 

\subsection{Emergent non-Abelian gauge field}
Within the approximation of intermediate wavevectors used above for the description of the spin-orbit coupling in polariton graphene, the Hamiltonian $\hat{H} = \hbar v_F \left(\hat{\sigma}_x\hat{k}_x+\hat{\sigma}_y\hat{k}_y \right)+\Delta\left(\hat{\sigma}_y\hat{s}_y -\hat{\sigma}_x\hat{s}_x\right)$ can be reformulated to become mathematically similar to that of a charged Dirac particle in presence of a vector potential of a gauge field. This is possible thanks to the reduced symmetry of the effective Dresselhaus spin-orbit coupling close to the Dirac point, as compared with the TE-TM field in the $\Gamma$ point \cite{Nalitov2014}. The Dirac Hamiltonian for a charged particle reads \cite{Dirac1930}:

\begin{equation}
\hat{H}=\hbar c \mathbf{\hat{\sigma}} \left(\mathbf{k}-\frac{e}{\hbar c}\mathbf{A}\right)
\end{equation}
where $\mathbf{A}$ is the vector potential of the electromagnetic field ($\mathbf{B}=\mathrm{rot} \mathbf{A}$). As for the Schrodinger equation, the momentum $\mathbf{p}$ is simply replaced by $\mathbf{p}-e/c\mathbf{A}$. Here, $e$ is the electron charge and $c$ is the speed of light, which in the case of the effective graphene Hamiltonian is replaced by the Fermi velocity $v_F$, determined by the coupling $J$ in the tight-binding model we consider.

To have the analogy with the Dirac equation, we need to define the effective vector potential as:

\begin{equation}
\mathbf{A}=-\frac{\Delta \mathbf{\hat{s'}}}{\hbar e}
\end{equation} 
where we have inverted one of the axes of the polarization pseudospin: $\hat{s'}_x=-\hat{s}_x$, $\hat{s'}_y=\hat{s}_y$. In the new coordinates of the polarization pseudospin space, the Dresselhaus field is converted into the Rashba field. Using this vector potential, the polariton graphene Hamiltonian can be written in the gauge field representation:

\begin{equation}
\hat{H}=\hbar v_F \mathbf{\hat{\sigma}} \left(\mathbf{k}-\frac{e}{\hbar v_F}\mathbf{A}\right)
\end{equation}
where the components of the vector potential $A_x$ and $A_y$ do not commute with each other. Thus, the gauge field is non-Abelian \cite{Hatano2007,Yang2008}. Written in this form, the equation for \emph{bispinor neutral} particles is reduced to a usual Dirac equation for \emph{spinor charged} particles in an emergent gauge field. Since the vector potential contains the polarization pseudospin operators $\hat{s}_{x,y}$, its spatial distribution depends on the current distribution of the polarization pseudospin, determined by the polarization spinor part of the wavefunction. Thus, the emergent electromagnetic field texture is defined by the polarization of polaritons. This leads to interesting effects, such as the lensing by an impenetrable defect, as was shown for other configurations with polaritonic emergent gauge fields \cite{Tercas2014}. Detailed study of the consequences of the emergence of the gauge field for Dirac equation is a subject for future works.

\section{Klein tunneling}
The description in terms of the Dirac Hamiltonian for excitations in honeycomb lattices is well established, and Klein tunneling in graphene \cite{Huard2007,Young2009}, as well as with atomic condensates in optical lattices \cite{Salger2011}, has already been demonstrated experimentally. Although this phenomenon is described in many review papers \cite{Calogeracos1998,Calogeracos1999,Beenakker2008,Pereira2010,Allain2011}, its mathematical description in the simple case is used as a basis for the spinor case considered in this manuscript. This is why we briefly revisit the simplest scalar case in the subsection below.

\subsection{Scalar case}
Qualitatively, Klein tunneling for massless particles consists in a perfect transmission through potential barriers because of the particle-antiparticle conversion possible for the Dirac equation: a particle with energy $E<V$ turns into an antiparticle with energy $E'=E-V$, propagating in the same direction, as shown in Fig. 1(a) by a green arrow (dispersion is shown in red). The backscattering is suppressed because of the pseudospin conservation: the potential cannot change the lattice pseudospin, because it acts identically on both components $A$ and $B$. Of course, this is true only for a particular propagation direction or in Born approximation.

For the general description of Klein tunneling, we need first to write the spinor solution of the Dirac equation \ref{scalarDirac} in the $A,B$ atom basis for an arbitrary propagation direction

\begin{equation}
\label{propag}
\Psi = \frac{1}{\sqrt{2}} e^{i \mathbf{k.r}} 
\begin{pmatrix}
1 \\ \alpha e^{i \phi}
\end{pmatrix}
\end{equation}

The propagation direction with respect to the horizontal axis is given by the angle $\phi$ with  $\tan(\phi) = k_y/k_x$ as shown on the Fig. 1(b). Let us now consider the incidence of a wave defined by Eq.\eqref{propag} on a potential barrier of a height $V$ located at $x=0$ (uniform along $y$).

The energy of the particle given by $E=\hbar v_F |k|$ in the Klein tunneling regime is smaller than the barrier height $V$.  The incident, reflected, and transmitted wavefunctions can be written as:

\begin{eqnarray}
 \Psi_i &=& \frac{1}{\sqrt{2}} e^{i \mathbf{k_i.r}} 
\begin{pmatrix}
1 \\ \alpha_i e^{i \phi_i}
\end{pmatrix}
\quad
\Psi_r = \frac{r}{\sqrt{2}} e^{i \mathbf{k_r .r}}
\begin{pmatrix}
1 \\ \alpha e^{i \phi_r}
\end{pmatrix}
\\
\Psi_t &=& \frac{t}{\sqrt{2}} e^{i \mathbf{k_t .r}}
\begin{pmatrix}
1 \\ \alpha' e^{i \phi_t}
\end{pmatrix} \nonumber
\end{eqnarray}
The continuity of the wavefunction in $x = 0$ imposes the following constrains on the two components of the spinor:

\begin{equation}
\left \{ 
\begin{array}{r c l} 
e^{i k^{i}_y y} + r e^{i k^{r}_y y} & = & t e^{i k^{t}_y y} \\
\alpha e^{i k^{i}_y y} e^{i \phi} + r \alpha e^{i k^{r}_y y} e^{i \phi_r} & = & \alpha' t e^{i k^{t}_y y} e^{i \phi_t}
\end{array}
\right.
\end{equation} 
The invariance in the $y$ direction imposes the conservation of the wave vector $ k^{i}_y = k^{r}_y = k^{t}_y$. Since $E<V$, $\alpha = 1$ and $\alpha' = -1$. This allows to determine the remaining unknowns: $\phi_r = \pi - \phi_i$ and $-(E-V_0)\sin( \phi_t) = E \sin( \phi_i )$. The equations become:

\begin{equation}
\left \{ 
\begin{array}{r c l} 
1 + r & = & t \\
e^{i \phi_i} - r e^{-i \phi_i} & = & - t e^{i \phi_t}
\end{array}
\right.
\end{equation}

Which leads to:

\begin{equation}
r = \frac{ e^{i \phi_i} + e^{i \phi_t}}{ e^{-i \phi_i} - e^{i \phi_t}} \ , \ \ \ t = \frac{2\cos( \phi_i )}{e^{-i \phi_i} - e^{i \phi_t}}
\end{equation}

The expressions become particularly simple for $ V = 2 E $, giving $E\sin(\phi_t)=E\sin(\phi_i)$, or simply $\phi_t=\phi_r=\pi-\phi_i$. The expressions for the reflection and tunneling amplitudes and intensities become:

\begin{equation}
\left \{ 
\begin{array}{r c l c r c l} 
r & = & e^{i( \phi_i + \pi / 2)} \sin( \phi_i ) \ \ & , & \ \ t & = & e^{i \phi_i} \cos( \phi_i ) \\
R & = & \sin^2 ( \phi_i ) \ \ & , & \ \ T & = & \cos^2 ( \phi_i )
\end{array}
\right.
\end{equation} 

One can see that indeed, for normal incidence $\phi_i=0$ the reflection is suppressed, and the transmission is $T=1$. This is the famous Klein tunneling effect, relying on the particular shape of the dispersion.

\subsection{Klein tunneling in the spinor case}

The extra degree of freedom given by the polarization of light allows to define a second pseudospin, independent from the pseudospin associated with the lattice sites $A$ and $B$. This second pseudospin corresponds to the Stockes vector of light, and it is related to the components of the spinor in the circular basis as:

\begin{eqnarray}
{S_{x}} &=&\Re \left( {\psi _{+}^{ph}\psi _{-}^{ph\ast }}\right)   \nonumber
\\
{S_{y}} &=&\Im \left( {\psi _{+}^{ph\ast }\psi _{-}^{ph}}\right)  \\
{S_{z}} &=&{{\left( {n_{+}^{ph}-n_{-}^{ph}}\right) }\mathord{\left/
{\vphantom {{\left( {n_ + ^{ph} - n_ - ^{ph}} \right)} 2}} \right.
\kern-\nulldelimiterspace}2}  \nonumber
\end{eqnarray}

In this section, we are going to consider the specific case of a potential barrier present only for one of the two spin components (say, $\sigma^+$), and absent for the other spin component ($\sigma^-$), which can be realized because of the spin-anisotropic interactions of polaritons and by using circularly polarized resonant or non-resonant pumping to create the potential barrier  \cite{Amo2010} (see Annex I). Since the two wavefunctions are uncoupled (we neglect the spin-orbit coupling in this section), the Klein tunneling (with the corresponding modification of the wave function) will occur only for one component, whereas the other will be just freely propagating. It is easy to write the solution of the Dirac equation in this particular case, combining the homogeneous solution for $\sigma^-$ with the solution exhibiting the particle-hole transition for $\sigma^+$. In the barrier region, the action of the spin-polarized potential is qualitatively similar to that of a magnetic field causing a Zeeman splitting between the circular components: the linear polarization will precess around this field.

%To this Hamiltonian, one should add a potential $V$, acting selectively only on one circular polarization component.

The wavefunction of a linearly polarized state propagating in a particular direction reads: $\Phi=\frac{1}{2} e^{i\mathrm{kr}}\left(1,\alpha e^{i\phi},e^{i\theta},e^{i(\phi+\theta)}\right)^{\mathrm{T}}$, where $\tan{\phi}=k_y/k_x$ as before, and $\theta$ is the relative phase which determines the orientation of linear polarization (or the direction of the pseudospin), while $\alpha$ gives the sign of energy (particles or holes). Since the two polarization components are essentially independent in the absence of spin-orbit coupling, it is useful to write the corresponding spinors separately: $\Psi^{+}=(\Psi_A^+,\Psi_B^+)^{\mathrm{T}}$, $\Psi^{-}=(\Psi_A^-,\Psi_B^-)^{\mathrm{T}}$. The incident, reflected and transmitted wave functions for both polarization components are given by the following expressions, based on the previous results for the reflection and transmission coefficients in the Klein tunneling regime for the $\sigma^+$ component (still assuming $E=V/2$ for simplicity; $k=|\mathbf{k}|$):

\begin{equation}
\left \{
\begin{array}{r c l}
\Psi^{+}_{i} & = & \frac{1}{ \sqrt{2}} e^{i k(x \cos \phi + y \sin \phi)} \begin{pmatrix}
1 \\ e^{i \phi}
\end{pmatrix} \\
\Psi^{-}_{i} & = & \Psi^{+}_{i} \\
\Psi^{+}_{r} & = & \frac{\sin \phi}{ \sqrt{2}} e^{i (k(-x \cos \phi + y \sin \phi) + \phi + \pi /2)} \begin{pmatrix}
1 \\ -e^{-i \phi}
\end{pmatrix} \\
\Psi^{-}_{r} & = & 0 \\
\Psi^{+}_{t} & = & \frac{\cos \phi}{ \sqrt{2}} e^{i (k(-x \cos \phi + y \sin \phi) + \phi)} \begin{pmatrix}
1 \\ e^{-i \phi}
\end{pmatrix} \\
\Psi^{-}_{t} & = & \Psi^{+}_{i} =  \frac{1}{ \sqrt{2}} e^{i k(x \cos \phi + y \sin \phi)} \begin{pmatrix}
1 \\ e^{i \phi}
\end{pmatrix}
\end{array}
\right. 
\end{equation}

There is no reflected wave for $\sigma^-$, because the barrier is present only for $\sigma^+$. Moreover, in the case of normal incidence, there is no reflection for $\sigma^+$ as well, as can be seen from the $\sin(\phi)$ factor in $\Psi_r^+$. In this case, the transmission for both polarizations is equally perfect, but a relative phase appears between them, as can be seen comparing $\Psi_t^+$ and $\Psi_t^-$: for $\phi=0$, $\Psi_t^+/\Psi_t^-=\exp(-2ikx)$. Since the relative phase between the spin components is what determines the orientation of the linear polarization, let us calculate it explicitly by analyzing the polarization pseudospin in the general case of $\phi\neq0$. We have:
\begin{equation}
\begin{array}{r c l}
 \Psi_L & = & \Psi_i + \Psi_r  \\
 \Psi_R & = & \Psi_t 
\end{array} 
\end{equation}

We obtain 8 pseudospin components : $x$ and $y$ on the $A$ and $B$ sub-lattices to the left and to the right of the barrier.
\begin{widetext}
\begin{equation}
\label{pres1}
\left \{
\begin{array}{r c l}
S^{L,A}_x & = & \frac{1}{2} ( 1 - \sin \phi \sin ( \phi - 2 k x \cos \phi )) \\ \\
S^{L,A}_y & = & - \frac{1}{2}  \sin \phi \cos ( \phi - 2 k x \cos \phi ) \\ \\
S^{L,B}_x & = & \frac{1}{2} ( 1 - \sin \phi \sin ( \phi + 2 k x \cos \phi )) \\ \\
S^{L,B}_y & = & \frac{1}{2}  \sin \phi \cos ( \phi + 2 k x \cos \phi ))
\end{array}
\right.
\ \ \ \ \left \{
\begin{array}{r c l}
S^{R,A}_x & = & \frac{1}{2}  \cos \phi \cos ( \phi - 2 k x \cos \phi ) \\ \\
S^{R,A}_y & = & - \frac{1}{2}  \cos \phi \sin ( \phi - 2 k x \cos \phi ) \\ \\
S^{R,B}_x & = & \frac{1}{2}  \cos \phi \cos ( \phi + 2 k x \cos \phi ) \\ \\
S^{R,B}_y & = & \frac{1}{2}  \cos \phi \sin ( \phi + 2 k x \cos \phi )
\end{array}
\right.
\end{equation}
\end{widetext}

One can easily see that these expressions verify the continuity at $x=0$. From them, one can plot the linear polarization degree in any basis (horizontal/vertical or diagonal/anti-diagonal) for any angle of incidence, or simply plot the orientation of the linear polarization plane $\eta=1/2\arctan S_y/S_x $. We note that the linear polarization for the two sublattices $A$ and $B$ is not the same: the diagonal polarization can be different for $\phi\neq 0$. The main consequence is the existence of spatial variations of the linear polarization on both sides of the spin-polarized barrier.

In the case of normal incidence, the expressions are strongly simplified:
\begin{equation}
\left \{
\begin{array}{r c c c l}
S^{L,A}_x & = & S^{L,B}_x & = & \frac{1}{2} \\ \\
S^{L,A}_y & = & S^{L,B}_y & = & 0
\end{array} 
\right.
\ \ \ \ \left \{
\begin{array}{r c c c l}
S^{R,A}_x & =  S^{R,B}_x & =  \frac{1}{2} \cos(2kx) \\ \\
S^{R,A}_y & =  S^{R,B}_y & =  \frac{1}{2} \sin(2kx)
\end{array}
\right.
\end{equation}
In this case, spatial oscillations of the linear polarization are observed only to the right of the barrier, in the transmitted wave.

Fig. 2 presents the results of the calculations of the pseudospin components according to the equations \eqref{pres1} for normal incidence in panel (a) and for oblique incidence in panel (b). One can see that the polarization always rotates in the region of the barrier, whereas in the left half-plane the rotation depends on the angle of incidence. Panel (a) also demonstrates the full transmission, signature of the Klein tunneling regime.

\begin{figure}[t]\label{fig2}
\includegraphics[scale=0.3]{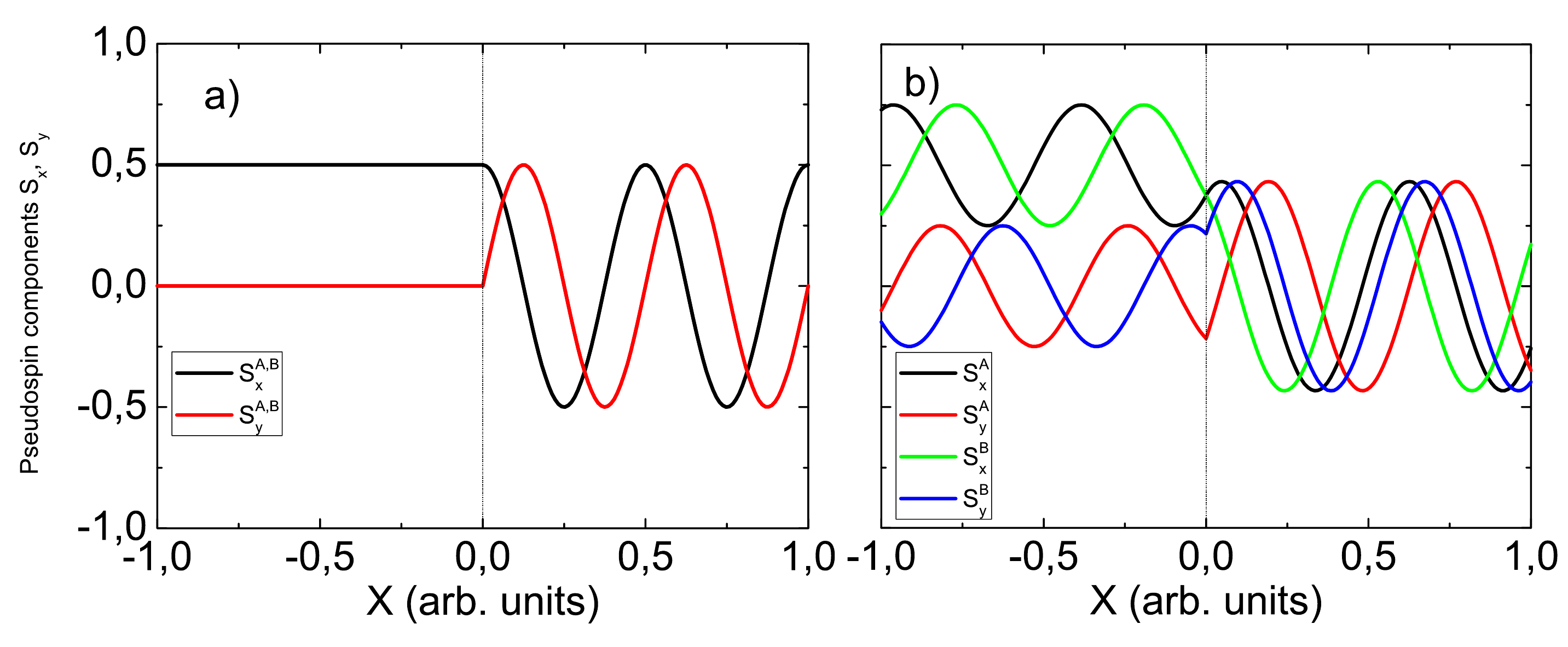} 
\caption{(color online) Pseudospin components for normal $\phi=0$ (panel a) and oblique $\phi=\pi/6$ (panel b) incidence. Linear polarization always rotates in the right half-plane (barrier region). }
\end{figure} 

The above description captures the important specificity of polariton graphene, that is, the presence of two different pseudospins, and demonstrates its consequences, but for the moment we were neglecting the spin-orbit interaction, which can play an important role for Klein tunneling, as shown previously for other systems \cite{Zhang2012}.

\subsection{Suppression of the Klein tunneling in presence of spin-orbit coupling}

In this section, we consider the same as previously, namely a potential barrier acting only on one spin component, but including the photonic spin-orbit coupling. The corresponding dispersion obtained in the section 1, parabolic at small wavevectors and linear at larger wavevectors, is shown in Fig. 3(a). The corresponding eigenvectors, numbered in the order of increasing energy for a fixed wavevector, are as follows:
\begin{widetext}
\begin{equation}
\Psi_1 = \begin{pmatrix}
-1 \\ \frac{\hbar v_F k}{E_{1}} e^{-i \phi} \\ -\frac{\hbar v_F k}{E_1} e^{i \phi} \\ 1
\end{pmatrix}, \quad
\Psi_2 =\begin{pmatrix}
1 \\ \frac{\hbar v_F k}{E_2} e^{-i \phi} \\ \frac{\hbar v_F k}{E_2} e^{i \phi} \\ 1
\end{pmatrix}, \quad
\Psi_3 =\begin{pmatrix}
-1 \\ \frac{\hbar v_F k}{E_3} e^{-i \phi} \\ -\frac{\hbar v_F k}{E_3} e^{i \phi} \\ 1
\end{pmatrix}, \quad
\Psi_4 =\begin{pmatrix}
1 \\ \frac{\hbar v_F k}{E_4} e^{-i \phi} \\ \frac{\hbar v_F k}{E_4} e^{i \phi} \\ 1
\end{pmatrix}.
\end{equation}
\end{widetext}

To find analytically the reflection and transmission coefficients we use the same approximations as before: the potential barrier is invariant in the $Y$ direction, the energy is one half of the barrier height $ E=V/2 $. Another assumption is linked with the choice of the initial energy branch, determining the eigenvector. One should take into account that transitions between the branches occurring in presence of a potential $V$ should conserve the polarization in the vicinity of the barrier edge, which allows only $3\rightarrow 1$ and $4\rightarrow 2$ transitions. 
 We assume that the incident wave corresponds to the 4th branch (positive and split-off). This gives us the following system of equations:

\begin{equation}
\left \{
\begin{array}{r c l}
1 + r & = & t \\
\frac{\hbar v_F k}{E} (e^{-i \phi} - r e^{i \phi}) & = & t \frac{\hbar v_F k_t}{E'} e^{-i \phi_t} \\
\frac{\hbar v_F k}{E} (e^{i \phi} - r e^{-i \phi}) & = & t \frac{\hbar v_F k_t}{E'} e^{i \phi_t}
\end{array}
\right.
\end{equation}

With this system we determine three parameters : $r$, $t$ and $ \phi_t $.

\begin{eqnarray}
r &=& \frac{k e^{i \phi} + k_t e^{i \phi_t}}{k e^{-i \phi} - k_t e^{i \phi_t}} \\
t &=& \frac{2k cos \phi}{k e^{-i \phi} - k_t e^{i \phi_t}} \\
\phi_t &=& \pi + \arcsin ( \frac{k}{k_t} sin \phi )
\end{eqnarray}

In the case of normal incidence, the expressions can be greatly simplified.
\begin{equation}
 r = \frac{k - k_t}{k + k_t}, \quad
 t = \frac{2k}{k + k_t} 
\end{equation}

We can thus determine the reflection and transmission probabilities $ R = |r|^2 $ and $ T = 1 - R $, and analyze how they are affected by the spin-orbit coupling $\Delta$. 
Since the Klein tunneling is associated with the linear dispersion, it is natural to expect the suppression of the transmission for the parabolic part of the dispersion. The results of the calculations are represented in Fig. 3(b) as a function of wavevector (plotted for $ \hbar = v_F = \Delta = 1 , \phi = 0 $ ). Within our assumptions, $k_t=\sqrt{4\Delta^2+4\Delta\sqrt{\Delta^2+k^2}+k^2}$.

\begin{figure}[tbp] 
   \centering \includegraphics[scale=0.35]{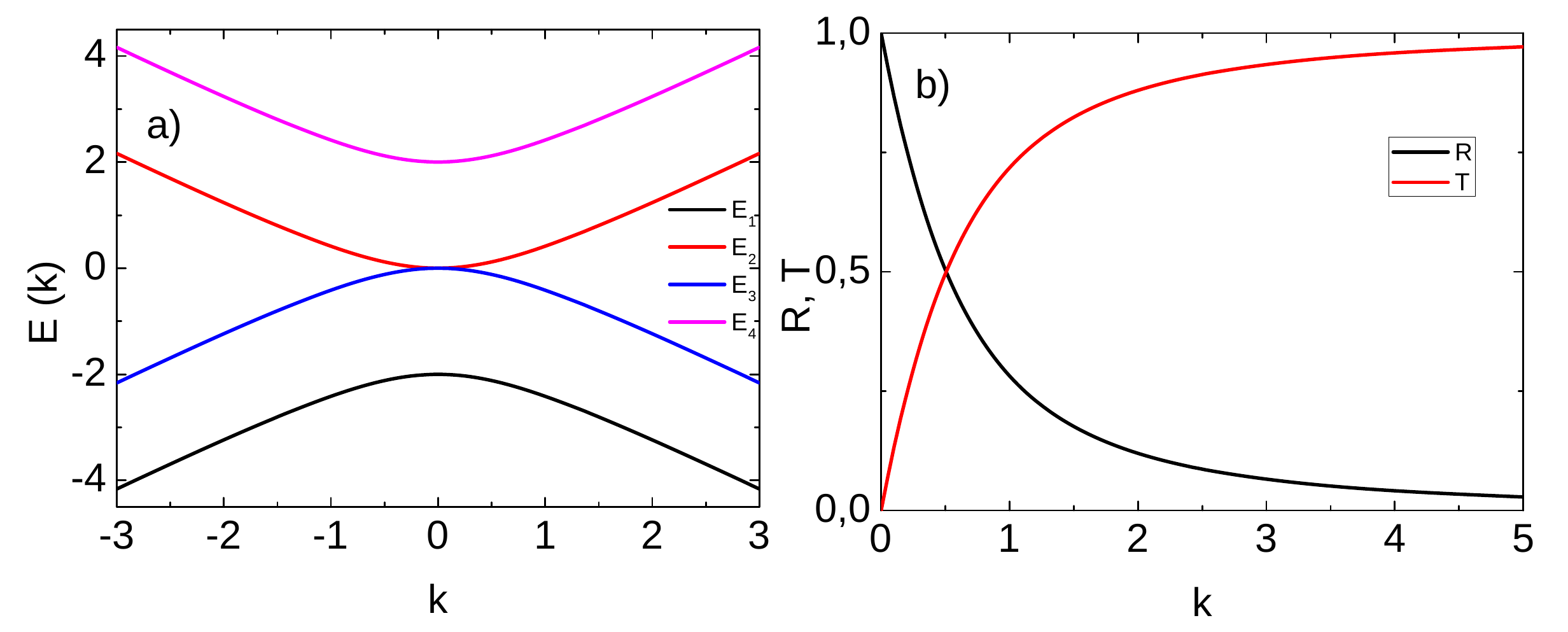}
   \caption{ (Color online) a) Dispersion in the vicinity of the Dirac point in presence of spin-orbit coupling. b) Transmission and reflection coefficients for normal incidence}
\end{figure}

 At lower wavevectors, the transmission becomes strongly suppressed, because the Klein mechanism does not protect it anymore. However, for higher wavevectors, where the dispersion is linear, the Klein tunneling is not suppressed, and therefore the results obtained in the previous subsection for decoupled spin components remain valid. In realistic structures, the energy difference between the split-off bands is comparable with the TE-TM splitting magnitude at the wave vector of the Dirac point. It depends completely of the structure geometry but typically ranges between a few tens and 100$~\mu$eV. It is therefore of the order of the mode linewidth in good quality samples. We therefore expect the Klein tunneling suppression to be an observable but relatively weak effect.
 
\section{Klein polarization rotator}
 
 The rotation of polarization on a micrometer scale in the absence of any backscattering (for normal incidence) allows to use the proposed structure as a polarization rotator, which can be called "Klein polarization rotator" or "Klein waveplate".   The deterministic control of polariton spin, associated with the other opportunities offered by polariton graphene structures, such as one-way surface states, makes of it a promising platform for spin-optronic applications.

To check our analytical predictions, we have performed a numerical simulation based on a spinor Schrodinger equation for polaritons where the honeycomb confining potential $U(x,y)$ is taken into account. We consider both the situation with and without spin-orbit coupling. Without spin-orbit coupling the equation reads:
\begin{eqnarray}
& i\hbar \frac{{\partial \psi _ \pm  }}
{{\partial t}}  =  - \frac{{\hbar ^2 }}
{{2m}}\Delta \psi _ \pm   + U\psi _ \pm   - \frac{{i\hbar }}
{{2\tau }}\psi _ \pm   + \\
& +P_0 e^{ { - \frac{{\left( {{\mathbf{r}} - {\mathbf{r}}_0 } \right)^2 }}
{{\sigma ^2 }}}}e^{ {i\left( {{\mathbf{kr}} - \omega t} \right)} } \notag
\end{eqnarray}
where $\psi(r)={\psi_+(r), \psi_-(r)}$ are the two circular components of the  wave function, $m$ is the polariton mass, $\tau=25$ ps the lifetime. Since the calculation is performed without the tight-binding approximation, only the polarization pseudospin remains. We have taken $m=5\times10^{-5}m_0$, where $m_0$ is the free electron mass.
$P_{0}$ is the amplitude of the pumping (identical for both components, corresponding to horizontal polarization), the size of the spot $\sigma=5$ $\mu$m in the $X$ direction and 40 $\mu$m in the $Y$ direction. The result of the simulation is shown in Fig. 4, demonstrating the inversion of the linear polarization degree just after the barrier (located at $x=0$), which appears in blue on the figure.

\begin{figure}[t]\label{fig2}
\includegraphics[scale=0.4]{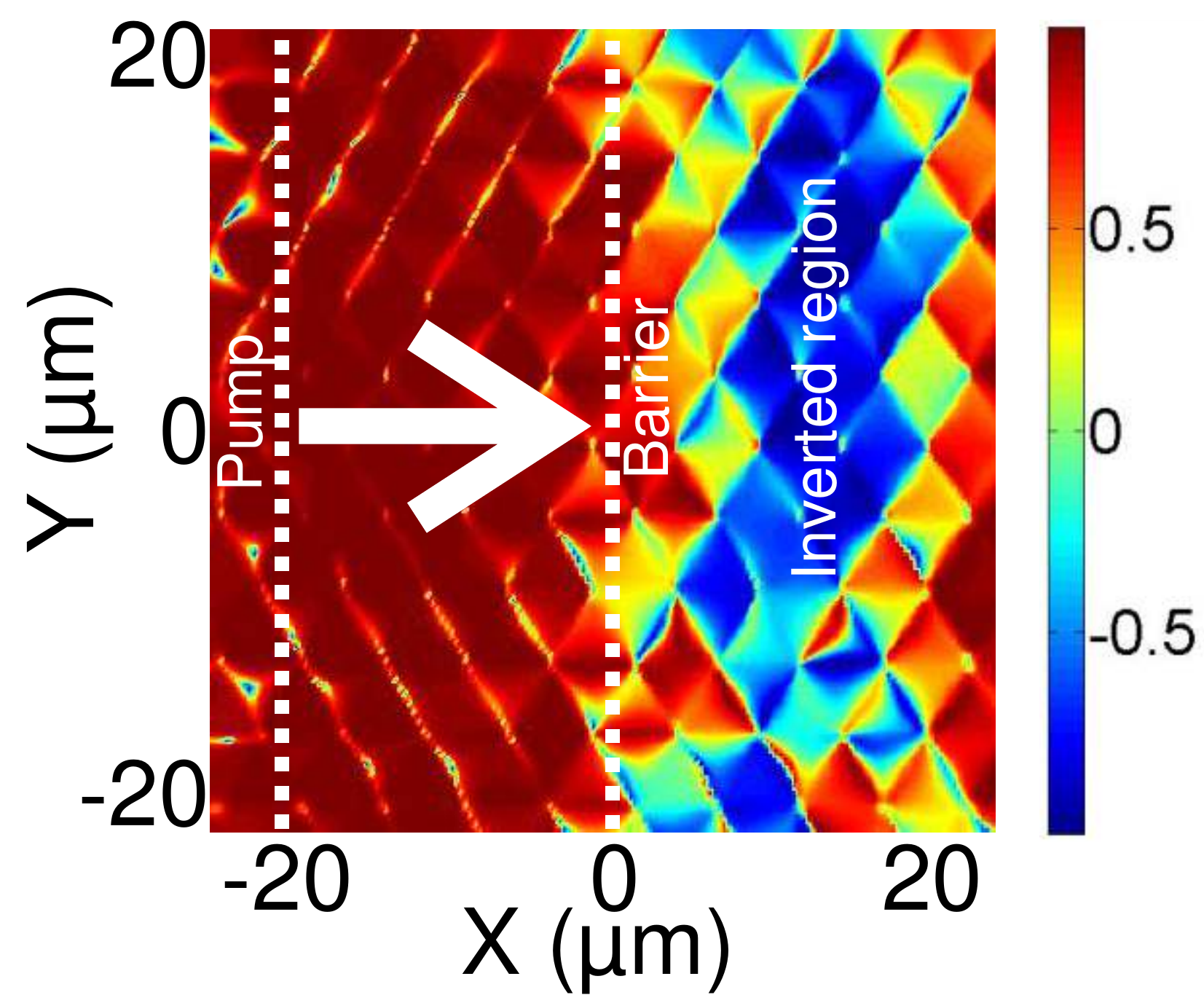} 
\caption{(color online) Linear polarization degree in a 2D polariton graphene. The pump is located at the left edge of the figure, polaritons propagate to the right. A barrier for $\sigma^+$ component is located at $X=0$, which leads to the inversion of the polarization degree at $X=10~\mu$m. Color shows linear polarization degree $\rho_l=(I_H-I_V)/(I_H+I_V)$. }
\end{figure} 

We see that the numerical simulations confirm the analytical predictions. Indeed, a barrier for one polarization component can be considered as an effective magnetic field in the $Z$ direction, because it creates a "Zeeman" energy splitting between circular polarization components. This effective field rotates the linear polarization in the plane, which is at the origin of the observed effect, while the Klein tunneling regime provides the suppression of backscattering. The average value of backscattered intensity, mostly due to the finite size of the beam leading to the violation of the normal incidence condition on the edges, did not exceed 1$\%$.

In order to check that the operation of the "Klein waveplate" remains possible in spite of the spin-orbit coupling existing in all real structures, we have repeated the numerical simulations, but with the new Hamiltonian and energy dispersion, adding the spin-orbit couping (without applying the linear approximation required for analytical calculations). For this, an extra term was added into the Hamiltonian of the spinor Schrodinger equation:
\begin{equation}
 \hat{H}\psi_\pm=\hat{H_0}\psi_\pm+\beta {\left( {\frac{\partial }{{\partial x}} \mp i\frac{\partial }{{\partial y}}} \right)^2}{\psi _ \mp } 
\end{equation}
The TE-TM splitting \cite{Shelykh2010} is described by the parameter $\beta ={%
\hbar ^{2}}\left( {m_{l}^{-1}-m_{t}^{-1}}\right) /4m$ where $%
m_{l,t}$ are the effective masses of TM and TE polarized particles
respectively and $m=2\left( {{m_{t}}-{m_{l}}}\right) /{m_{t}}{%
m_{l}}$. The results of the simulations are presented in Fig. 5 (a) and (b), the difference between them being the value of $\beta$. For Fig. 5(a) we have taken the typical value of 5\% difference between the masses, whereas for Fig. 5(b) this value was multiplied by a factor 5, corresponding to an artificially high value of TE-TM splitting, never achieved in real cavities.

\begin{figure}[tbp] 
   \centering \includegraphics[scale=0.35]{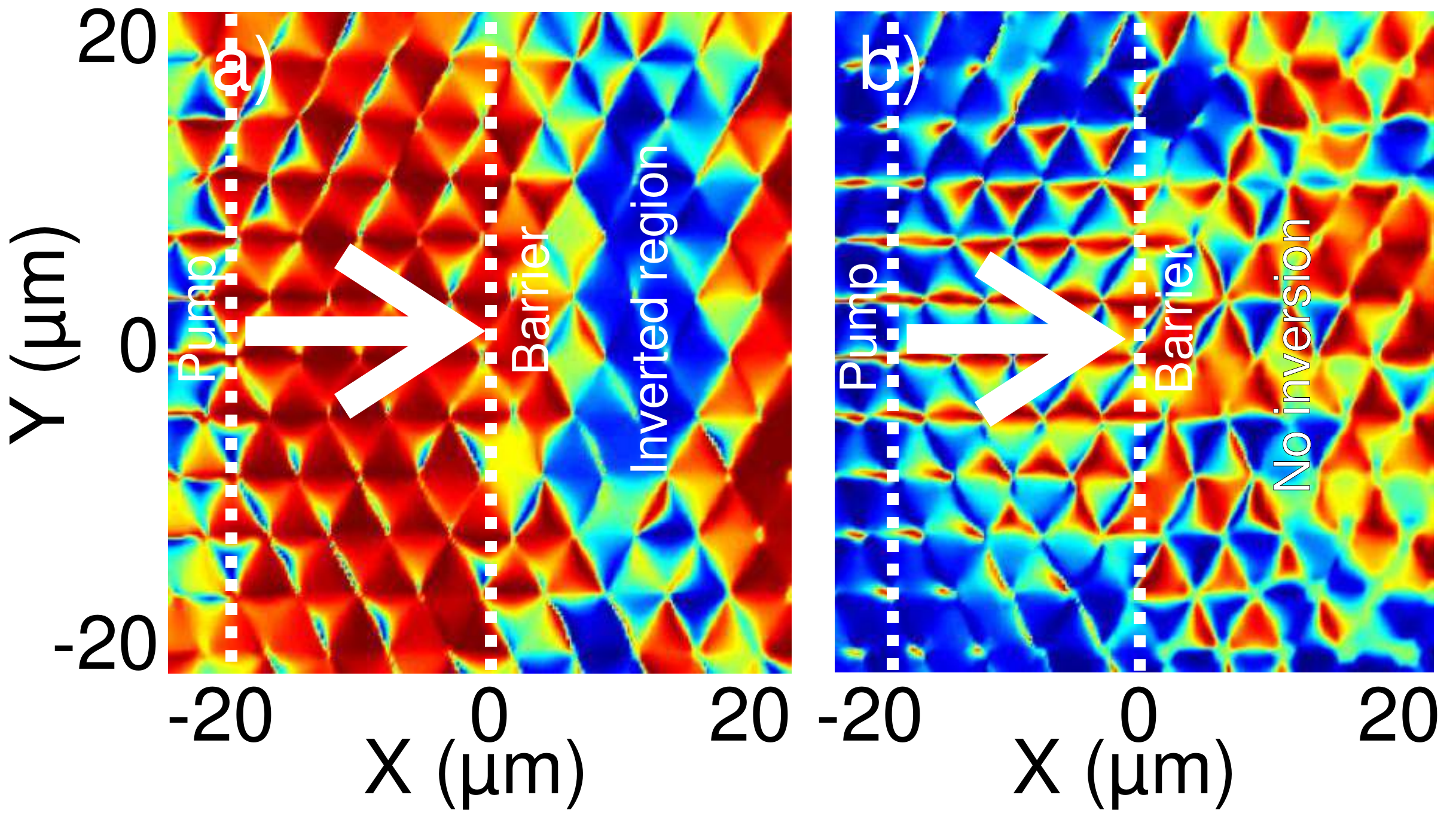}
      \caption{ (color online) Spatial images calculated by solving the full spinor Schrodinger equation for the "Klein waveplate" with TE-TM splitting, showing inear polarization degree with a) normal TE-TM splitting, b) TE-TM splitting $\times 5$. Color shows linear polarization degree (same as fig.3). }
\end{figure}

Once again the numerical simulations confirm the analytical calculations. The typical spin-orbit coupling existing in polariton graphene is too small to create any pronounced effects in the configuration of our simulated experiment: the "Klein waveplate" on Fig. 5(a) operates as expected, in spite of the non-zero value of $\beta$ and small parabolicity of the branches: a region of inverted polarization (blue) appears after the barrier. We did not observe any significant increase of the backscattered intensity (1\%) with respect to the case without spin-orbit coupling (Fig. 4). On the other hand, an artificially enhanced value of the spin-orbit coupling $\beta$ leads to the suppression of Klein tunneling (20\% of backscattering of the circular component with barrier) and the associated polarization rotation, as we can see in Fig. 5(b): there is no region with pronounced inversion of polarization in the right part of the figure. Furthermore, the linear polarization degree is not positive in the left part any more, which is a signature of a strong reflection on the barrier, together with polarization inversion.

\section{Conclusions}
To conclude, we have studied the Klein tunneling in polariton graphene, taking into account the spinor properties of polaritons, including the spin-anisotropic interactions and the spin-orbit coupling. We have shown that while the interactions allow to exploit the Klein tunneling effect for the creation of a micron size polarization rotator without reflection on its surfaces, the spin-orbit coupling might perturb its operation. However, the effect of this coupling for realistic structures remains negligible.

We acknowledge the support of ANR GANEX (207681) and Quandyde (ANR-11-BS10-001), and EU ITN INDEX (289968) projects.

\section{Annex I}

Polaritons, due to their mixed light-matter nature, exhibit an interesting property, leading to many curious effects: strongly spin-anisotropic interactions. Indeed, the interaction in the triplet configuration (same spins) is based on the exchange mechanism, and its strength is almost the same as for bare excitons, whereas in the singlet configuration an exchange leads to the formation of a dark exciton, whose energy is much higher than that of a lower polariton. It is therefore a second-order mechanism, which is strongly suppressed. This concerns not only interaction between two polaritons, but also between a polariton and a reservoir exciton. In both cases, the conclusion is that different potentials can be created for the two spin components using circular polarized optical pumping, either resonant (creating polaritons) or non-resonant (creating reservoir excitons). 

\bibliography{reference}

\end{document}